\newcommand{\tr}[1]{\mathrm{Tr}\left[ {#1} \right]} 

\newcommand{\ket}[1]{\vert #1\rangle}

\documentclass[twocolumn,pra,superscriptaddress]{revtex4-1}
\usepackage{bm,natbib}
\usepackage{inputenc}
\usepackage{graphicx}
\usepackage{mathrsfs}
\usepackage{dcolumn,fancyhdr}
\usepackage{amsmath}
\usepackage{amssymb}
\usepackage{amsfonts,gensymb}
\usepackage{indentfirst}
\usepackage{bbold}
\usepackage{multirow}
\usepackage{dsfont}
\usepackage{color}      

\begin{document}
\title{Detecting Gaussian entanglement via extractable work}

\author{Matteo Brunelli}
\affiliation{Centre for Theoretical Atomic, Molecular and Optical Physics, School of Mathematics and Physics, Queen's 
University, Belfast BT7 1NN, United Kingdom}
\author{Marco G. Genoni}
\affiliation{Quantum Technology Lab, Dipartimento di Fisica, Universit\`a degli Studi di Milano, 20133 Milano, Italy}
\author{Marco Barbieri}
\affiliation{Dipartimento di Scienze, Universit\`a degli Studi Roma Tre, Via della Vasca Navale 84, 00146, Rome, Italy}
\author{Mauro Paternostro}
\affiliation{Centre for Theoretical Atomic, Molecular and Optical Physics, School of Mathematics and Physics, Queen's 
University, Belfast BT7 1NN, United Kingdom}

\begin{abstract}

We show how the presence of entanglement in a bipartite Gaussian state can be detected by the amount of work extracted by a 
continuos variable Szilard-like device, where the bipartite state serves as the working medium of the engine. We provide an 
expression for the work extracted in such a process and specialize it to the case of Gaussian states.
The extractable work provides a sufficient condition to witness entanglement in generic two-mode states, becoming also necessary 
for squeezed thermal states. We extend the protocol to tripartite Gaussian states, and show that the full structure of inseparability 
classes cannot be discriminated based on the extractable work. This suggests that bipartite entanglement is the fundamental 
resource underpinning work extraction.

\end{abstract} 
\maketitle

One of the most striking -- to the point of being considered paradoxical for a long time -- and yet fundamental ways to extract work with the 
help of a heat engine, is to exploit the availability of information about the state of the engine itself. A machine following this paradigm 
is referred to as an information engine. In this way thermodynamics accommodates information in an operational way: 
the information acquired about a system effectively brings it out of equilibrium and useful work can be extracted by implementing suitable 
conditional operations~\cite{DemonRev, ThermoInfoRev}.   
\par
Recently there has been a lot of interest in exploring information-to-work conversion when the information is encoded in a quantum system~\cite{QuantumThermoRev}. 
For instance, fundamental thought experiments such as Maxwell's demons and Szil\'ard engine, have been formulated for quantum systems~\cite{MaxwellDemonLloyd,
MaxwellDemonZurek,QuantumSzilard}. Concerning work extraction, the most significative departure from a classical picture may be expected  
when the information is encoded in the correlations between two or more parties, in virtue of the unique role played by entanglement~\cite{EntanglementRev}. This has 
triggered  the study of work extraction from correlated quantum systems~\cite{ZurekDemon, WorkOppe,WorkFromCorr,GianlucaWork}. 
Yey little is known when such correlations are shared across a multipartite quantum working medium. 
\par
Interestingly, in Refs.~\cite{WorkVedral, WorkVedralTripartite} an alternative viewpoint was adopted by somehow reversing the question: what can the extractable work tell us about the nature 
of the correlations present in the working medium? Could it be used to discriminate  a separable state from an entangled one?
In the present work we build on this approach, using the extractable work as an investigative tool to gather some knowledge about the properties of a continuous-variable Szilard engine. 
We show how the extractable work is related to the one-way classical correlations established between two parties via a local measurement \cite{HendersonVedral}, and that 
it is a suitable quantity to witness bipartite entanglement in two mode Gaussian states~\cite{GaussRev, MatteoRev}.
 We further apply our diagnostics to tripartite Gaussian states, revealing how the work-extraction criterion overlooks differences in the inseparability classes. 
\par
We start by recalling the paradigm of Szil\'ard engine and information-to-work conversion in  Section~\ref{s:Szilard}. In Section~\ref{s:MultiSzilard} we formulate the work extracting protocol for 
correlated  quantum systems. In particular, in Sec.~\ref{s:OneMeas} and~\ref{s:TwoMeas} we address the relevant case of Gaussian states subjected to Gaussian measurements, and show
our main findings. An extension of the protocol beyond the Gaussian realm is discussed in Sec.~\ref{s:BeyondGauss}, while in Sec.~\ref{s:Tripartite} we attack the richer problem of work 
extraction from tripartite states. Finally, Sec.~\ref{s:Conclusions} reports our conclusions and some future perspective.

\section{Information-to-work conversion in a Szil\'ard engine}\label{s:Szilard}
In 1929 L\'eo Szil\'ard proposed a thought experiment, which now goes under the name of Szil\'ard engine, to highlight the link between information and thermodynamics 
and its apparently paradoxical consequences~\cite{Szilard}. Inspired by Maxwell's demon, he conceived a minimalist model to show how, through the acquisition of information 
and the implementation of feedback operations, the second law of thermodynamics may apparently be circumvented. Consider a  single particle in a box
with a frictionless wall that can be inserted and removed at half the length. If some information about the location of the particle becomes available, 
it can be exploited to extract some work (out of a freely available thermal bath) as follows: if the particle is known to be in one side of the container we can attach a weight on that side, in such a 
way that when we let the particle expands isothermally, the  ``pressure" exerted on the wall can pull up the weight. 
Assuming an isothermal expansion from the initial volume $V/2$ to the full volume we have $W=k_B T \ln 2$. After the 
expansion the system has returned to its initial configuration, so that the work extraction process can in principle be implemented cyclically. If the knowledge about the position of the 
particle is probabilistic, we have  
\begin{equation}\label{WSingle}
W=k_B T \ln 2 [1-H(X)]\, ,
\end{equation}
where $H(X)=-\sum_x p_x \ln p_x$, $x=\{R,L\}$, is the Shannon entropy of the right/left distribution. When both sides have the same probability the average extractable 
work is zero.    
This a priori information is usually symbolized by a demon, whose knowledge of the microscopic state of the system can be converted into useful work. 
Thanks to the demon's action, a thermodynamic cycle which generates work absorbing heat from a single reservoir may be realized.  
The paradoxical consequences of this thought experiment have attracted attention for quite a while, until Landauer recognized in the role of the memory the solution to the 
paradox~\cite{Landauer}. The demon needs to store the result of the 
measurements in a memory, and given that no physical memory can be taken to be infinite, the demon eventually needs to reset it in order to prevent overflow~\cite{Bennett}.
The erasure step is intrinsically irreversible and dissipates an amount of heat at lest equal to the work extracted in Eq.~\eqref{WorkExt}, thus restoring the second law.
Maxwell's demon-like devices and Landauer's erasure have been respectively realized and confirmed experimentally in recent years~\cite{Exp1,Exp2,Exp3,Exp4,Exp5}.

\section{Extracting work from correlated Szil\'ard engines}\label{s:MultiSzilard}
Imagine now to have two correlated particles, and suppose to trap them into separate containers, so to have two Szilard engines with correlated working substances, 
$A$ and $B$. The work extractable by one party, say $A$, now depends on the state of the other one, namely $W(A\vert B)=k_B T\log[1-H(A\vert B)]$. If some operation 
is performed on $B$ the state of knowledge of A must be updated. Given that the mutual information $I(A:B)=H(A)-H(A\vert B)\ge 0$ is nonnegative, we have 
$H(A)\ge H(A\vert B)$, that is conditioning reduces the uncertainty. It immediately follows
that  $W(A\vert B)\ge W(A)$ which proves that we can extract more work from correlated Szilard engines.
\par 
\begin{figure}[t!] 
\includegraphics[width=0.9\columnwidth]{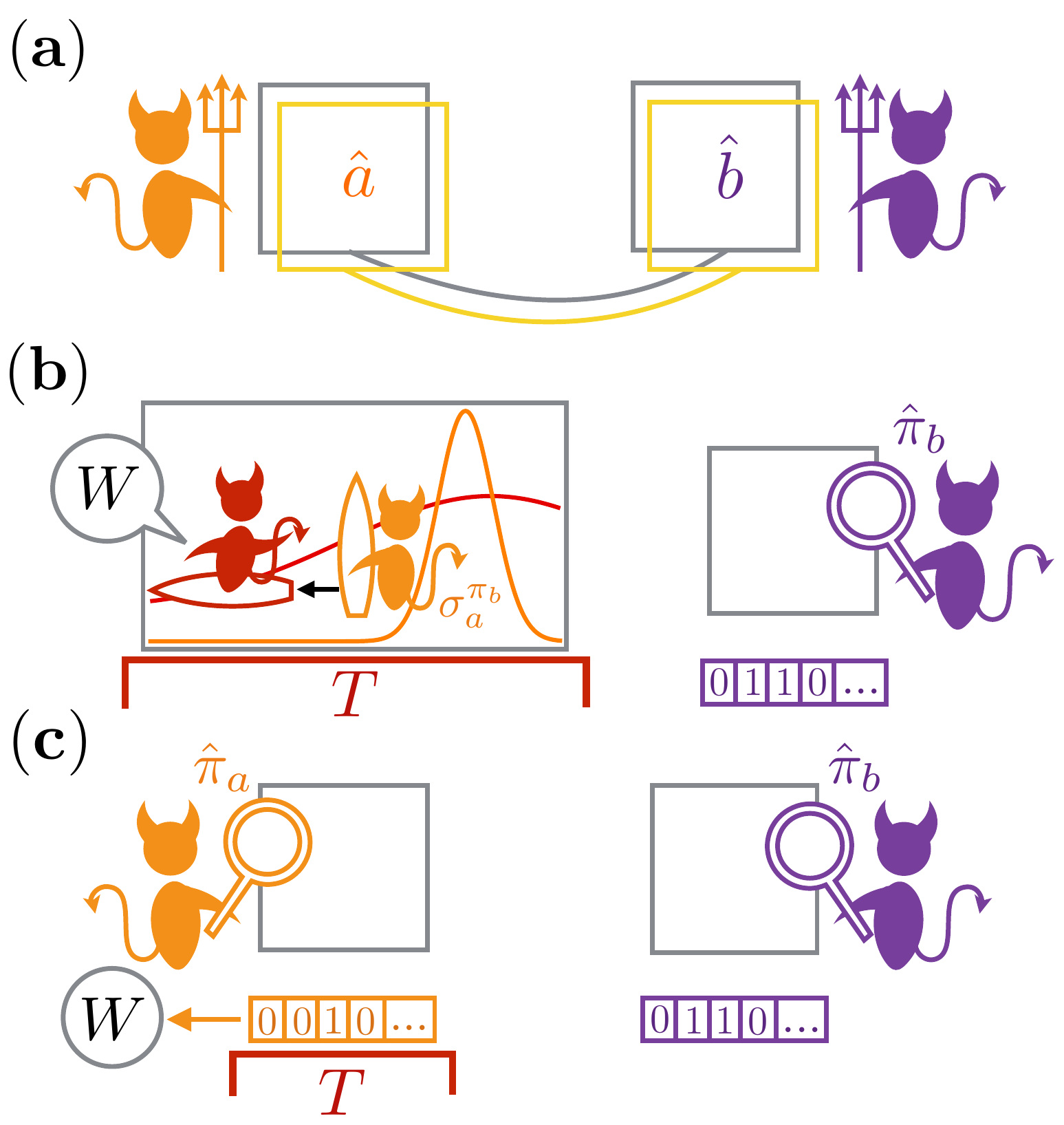}
\caption{({\bf a}) Gaussian demons Alice (orange) and Bob (purple) share a bipartite Gaussian state of modes $\hat a$ and $\hat b$ and want to know wether the state 
is entangled (yellow line) or separable (grey line). In order to do so, they check how much work Alice can extract from a heat bath when only local 
Gaussian measurements are allowed. In the first strategy ({\bf b}) Bob performs a Gaussian measurement $\hat\pi_b$ and Alice extract mechanical work by letting her 
conditional state $\sigma_a^{\pi_b}$ expand  (from orange to red), e.g. pushing the demon's board. As a result of the protocol Alice extracts an amount of work $W$. In 
a second approach ({\bf c}), both demons perform a measurement, and the work is extracted from the classical register of the results.   
\label{f:Protocol}}
\end{figure}

How can we extend this argument to quantum systems? 
In Ref.~\cite{WorkVedral} the authors considered Alice and Bob to share a bipartite system and perform projective measurements on {\it both} parties. 
The work is subsequently extracted by one party, say Alice, from the outcomes of the measurements, with Bob sharing with Alice his outcomes. 
In this context work extraction is to be understood as follows: each bit of information of the measurement outcome can be regarded as a particle in the left/right side of a container
(in principle the information can be copied in such a ``Szil\'ard register" without extra energy cost)~\cite{DemonRev}. In this way the work-extracting protocol 
is implemented at the level of the classical information obtained from a correlated quantum state via local
measurements and classical communication. This is reminiscent of a Bell-like scenario for testing local realism. 
In Ref.~\cite{WorkVedral} it was also shown that for separable states the work so extracted cannot exceed a limiting value, thus leading to a form of 
`work-assisted entanglement detection'. This protocol has been recently implemented in a photonic platform~\cite{WorkExp}.
\par
We generalize such an approach and study the inseparability of bipartite continuous-variable states by inspecting the amount of work extracted
by two local agents, or demons, Alice and Bob [see Fig.~\ref{f:Protocol} (\textbf{a})]. 
Let us notice that, in order to run an information engine, Alice does not need  to perform a measurement on her system and 
extract work from the recorded outcomes. She can exploit the back-action induced by Bob's measurement on their joint state and simply act locally by letting 
her state thermalize. The expansion can be converted into mechanical work. This work-extracting procedure is sketched in Fig.~\ref{f:Protocol} (\textbf{b})
with explicit reference to the Gaussian scenario, and discussed in the next section.
When both demons perform a local measurement, as in Fig.~\ref{f:Protocol} (\textbf{c}), the work is extracted by Alice from the register of measurements outcomes. 
\par

In the argument above, we have not considered the energetic and entropic cost of implementing the measurement. While this is certainly an important point to 
consider when attempting at investigating the thermodynamic balance associated to a given protocol, here the main scope is to use the extractable work as a
diagnostic tool to investigate the nature of quantum correlations. Therefore, the quantification of such costs is not crucial for our purposes.
For discussions on these issues see, e.g. Refs~\cite{Cost1,Cost2,Cost3}.   

\section{work extraction from bipartite gaussian states}\label{s:OneMeas}
In this section we explicitly formulate the work extracting protocol for Gaussian states sketched in Fig.~\ref{f:Protocol} (\textbf{b}) and
discuss the results. Gaussian demons Alice (orange) and Bob (purple) share a bipartite Gaussian state of modes $\hat a$ and $\hat b$
which is completely characterized by the covariance matrix
\begin{equation}\label{sigma}
\sigma_{ab}=
\left(
\begin{array}{cc}
\sigma_a \, & c_{ab} \\[1ex]
c_{ab}^T &  \sigma_b
\end{array}
\right) \, ,
\end{equation}
where $\sigma_{a (b)}$ is the reduced covariance matrix of Alice (Bob) while $c_{ab}$ contains the correlations between the modes. The first moments are inconsequential 
for our reasoning and can be set to zero. When a Gaussian state is in standard form we have~\cite{MatteoRev}
\begin{equation}\label{StandardForm}
\sigma_a=\text{diag}(a,a) \,, \quad \sigma_b=\text{diag}(b,b) \,, \quad c_{ab}=\text{diag}(c,d) \,.
\end{equation}
Bob performs a measurement on his mode. We restrict to Gaussian measurements of the form 
$\hat\pi_b(X)=\pi^{-1}\hat D_b (X)\hat{\varrho}^{\pi_b}\hat D_b^{\dagger}(X)$ where $\hat D_b(X)=\exp(X \hat b^{\dagger}-X^* \hat b)$ is the 
displacement operator, $\hat{\varrho}^{\pi_b}$ a pure Gaussian state with covariance matrix $\gamma^{\pi_b}=R(\phi)\text{diag}(\lambda/2,\lambda^{-1}/2)R(\phi)^T$, 
where $\lambda \in[0,\infty]$ and $R(\phi)=\cos\phi \mathbb{1}-i\sin\phi \sigma_y$ is a rotation matrix ($\sigma_y$ refers to the $y$-Pauli matrix). 
The conditional state of mode $\hat a$ on the measurement 
$\hat\pi_b(X)$ turns out to be independent on the outcome of the measurement itself, i.e. $\sigma^{\pi_b}_{a\vert X}\equiv \sigma_a^{\pi_b}$, and its expression is given by 
\begin{equation}\label{ReducedState}
\sigma_a^{\pi_b}=\sigma_a- c_{ab}(\sigma_b+\gamma^{\pi_b})^{-1}c_{ab}^T \, .
\end{equation} 
As a result of the measurement, the reduced state of mode $\hat a$ is now out of equilibrium 
and Alice can extract work from a heat bath by letting her state diffuse quasi-statically in the phase space [e.g., by pushing the board in Fig.~\ref{f:Protocol} (\textbf{b})]. 
She puts the system prepared in the post-measurement state in contact with the thermal bath and wait for it to reach equilibrium $\sigma_a^{\text{eq}}$.
As her state is independent on the outcome, its average entropy is simply $\int\text{d}X p_X S(\sigma_{a\vert X}^{\pi_b})=S(\sigma_a^{\pi_b})$. Following Eq.~\eqref{WSingle}
we can thus define the extractable work as
\begin{equation}\label{WorkExt}
W=k_B T \left[ S(\sigma_a^{\text{eq}})-S(\sigma_a^{\pi_b})\right] 
\end{equation}
Let us first address the simplest case in which $\sigma_{ab}$ is in the standard form~\eqref{StandardForm} and the reference thermal state has the same energy as Alice's initial 
state, i.e.  $\sigma_a^{\text{eq}}=\sigma_a$. In this way \textit{all the work extracted is due to measurement back-action}.
Indeed, we notice that the extractable work corresponds, up to a multiplicative factor, to the one-way classical correlations $\mathcal{J}^{\leftarrow}(\varrho_{ab})$, operationally 
associated with the distillable common randomness between the two parties \cite{DevetakWinter}. By maximizing it over all the possible measurements, it quantifies the total 
classical correlations between the two parties \cite{HendersonVedral}, and can be analytically evaluated for Gaussian states and Gaussian measurements \cite{GerardoDiscord,MatteoDiscord}.
\par
In order to quantify the entropy of the reduced state Eq.~\eqref{ReducedState}, we employ the R\'enyi entropy of order 2  $S_2(\varrho)=-\ln \tr{\varrho^2}$. 
When restricting to Gaussian states $S_2(\varrho)$ becomes a fully legitimate entropy functional, satisfying strong subadditivity~\cite{Adesso}, and takes a simple expression in terms of the
covariance matrix 
\begin{equation}\label{ShanWig}
S_2(\sigma_{ab})=\frac12 \ln (\det \sigma_{ab}) \, .
\end{equation}
The expression of the work Eq.~\eqref{WorkExt} then becomes
\begin{equation}\label{WorkExtRenyi}
W=\frac{k_B T}{2}\ln \left(\frac{\det \sigma_a}{\det\sigma_a^{\pi_b}}\right) \, .
\end{equation}
From now on we express the extractable work in units of $k_B T$. 
We recall that for our scope $W$ must be regarded the output of a suitable work-extraction protocol (that we consider as a black-box process). A non zero $W$, clearly corresponds to the presence of 
(classical) correlations between the two demons Alice and Bob. We will see that the knowledge of $W$, together with that of the local energies, 
always provides a sufficient criterion to detect entanglement. 
\par

\subsection{Symmetric squeezed thermal state}
\begin{figure*}[t!]
\includegraphics[width=2
\columnwidth]{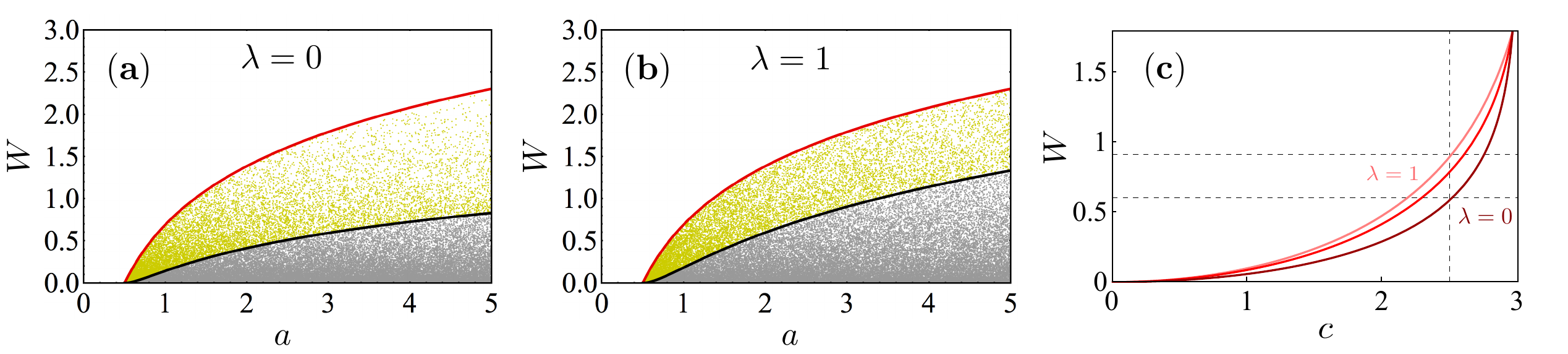}
\caption{Extractable work $W$ (in units of $k_BT$) against $a$ for randomly generated states. Each point corresponds to a state 
obtained by a uniform sampling of the parameters $a$ and $c$. points corresponding to entangled (separable) states are marked 
in yellow (gray). Panel ({\bf a}) refers to homodyne detection and panel ({\bf b}) to heterodyne detection. The red curve represents 
the maximum amount of extractable work $W_{\text{max}}$, while the black curve stands for the work at separability threshold 
$W_{\text{sep}}^{(k)}$, $k=0,1$. ({\bf c}) Extractable work  against the parameter $c$ for different Gaussian measurements and $a=3$. 
From lightest to darkest red: $\lambda=1,\, 5,\,0$.  
The vertical dashed lines refers to the value $c_{\text{sep}}=a-1/2$, while the horizontal ones to the corresponding values of 
$W_{\text{sep}}^{(k)}$, $k=0,1$.
\label{f:PlotWorkSymm}}
\end{figure*}

Let first address the case of quantum states of the form $\varrho_{ab} = S_2(r) \nu_\beta \otimes \nu_\beta S_2(r)^{\dag}$, generated by acting with a two-mode 
squeezing operator $S_2(r) = \exp\{ r (a^\dag b^\dag - a b)\}$ on two thermal states $\nu_\beta = e^{-\beta a^\dag a}/Z$ with the same temperature. Their corresponding 
covariance matrix is in standard form with $\sigma_a=\sigma_b=\text{diag}(a,a)$ and $c_{ab}=\text{diag}(c,-c)$, where $a\ge \frac12$ and $\vert c\vert \le \sqrt{a^2-\frac{1}{4}}$. 
Following Bob's measurement (with strength $\lambda$ and angle $\phi$), Alice can extract an average amount of work  given by
\begin{equation}\label{Wsymm}
W^{(\lambda)}=\frac{1}{2} \sum_{k=0,1}\ln \left[\frac{a (2 a\lambda^k +\lambda^{1-k})}{2( a^2-c^2) \lambda^k+a\lambda^{1-k}}\right] \, .
\end{equation} 
We notice there is no dependence on the measurement angle. In the limit $c\rightarrow0$ the expression vanishes, i.e. no work can be extracted from 
uncorrelated states. 
One can check that both the entanglement and $W^{(\lambda)}$ are monotonically increasing with the parameter $c$, and decreasing with the local energy 
parameter $a$. As a consequence, the maximum amount of work $W_{\text{sep}}^{(\lambda)}$ extractable by a separable state
is achieved at the separability threshold, which is given by $c_{\text{sep}}=a-1/2$. The condition $W^{(\lambda)}> W_{\text{sep}}^{(\lambda)}$ is therefore both 
necessary and sufficient for entanglement of $\sigma_{ab}$. The corresponding value of $W_\text{sep}^{(\lambda)}$ reads
\begin{equation}\label{WSepSymm}
W_{\text{sep}}^{(\lambda)}=\frac{1}{2}\sum_{k=0,1} \ln \left[\frac{2 a(2 a\lambda^k +\lambda^{1-k})}{( 4a-1) \lambda^k+2a\lambda^{1-k}}\right] \, .
\end{equation} 
Moreover, when the correlations attain the maximum value $c_{\text{max}}=\sqrt{a^2-1/4}$ (corresponding to a two-mode squeezed vacuum) the 
expression of the work is
\begin{equation}\label{WMaxSymm} 
W_{\text{max}}= \ln 2a\, ,
\end{equation} 
independently on the strength of the measurement.

In Fig.~\ref{f:PlotWorkSymm} (\textbf{a}), (\textbf{b}) we plot the curves Eqs.~\eqref{WSepSymm}, \eqref{WMaxSymm}  for the relevant case $\lambda=0$ 
($\lambda=1$) corresponding to homodyne (heterodyne) detection, together with randomly generated symmetric states. 
As expected, points corresponding to separable (gray) and entangled (yellow) states occupy disjoint regions, confirming how the extractable work provides 
a necessary and sufficient condition for separability. 
 From the plots it also is possible to see that for heterodyne measurements the maximum amount of work extractable from a separable state is 
larger than for the case of homodyne measurements. 
It is important to stress that the threshold is not universal, i.e. a constant value, but instead depends on the value of local energy: the couple $(a,W)$
then fully characterize the separability of the state. 
\par
Explicit expressions for $\lambda=0,1$ are listed below
\begin{equation}
W^{(0)}=\frac{1}{2} \ln \left(\frac{a^2}{a^2-c^2}\right)\,, \quad
W_{\text{sep}}^{(0)}=\frac{1}{2} \ln \left(\frac{4 a^2}{4 a-1}\right)\, \nonumber,
\end{equation}
and
\begin{equation}
W^{(1)}=\ln \left[\frac{a (2 a+1)}{2 (a^2- c^2)+a}\right]\,, \quad 
W_{\text{sep}}^{(1)}=\ln \left[\frac{2 a (2 a+1)}{1-6 a}\right]\,. \nonumber
\end{equation}

It is also instructive to look at the behavior of the extractable work $W$ against the correlations between the two modes. 
In Fig.~\ref{f:PlotWorkSymm} (\textbf{c}) we show the behavior of $W$ as a function of the parameter $c$ for a fixed value of the energy (fixed $a$). 
$W$ is monotonically increasing with respect to the amount correlations shared between the two modes. For product states ($c=0$) the
extractable work vanishes, while it achieves its maximum for a two-mode squeezed vacuum ($c=c_{\text{max}}$). Moreover, we can see that 
different measurement strategies allow for the extraction of different amounts of work. In particular we notice that the average work $W^{(\lambda)}$ 
extractable by implementing a Gaussian measurement  of strength $\lambda$ is both upper and lower bounded, i.e., $W^{(0)}\le W^{(\lambda)}\le W^{(1)}$. 
In particular heterodyne detection turns out to be optimal for work extraction. For any $a$ and $c$, $W^{(\lambda)}$ is monotonically increasing 
with respect to $\lambda$ in the interval $\lambda\in [0,1]$ and monotonically decreasing in $\lambda\in [1,\infty)$. 

\subsection{Squeezed thermal state}
\begin{figure}[t!] 
\includegraphics[scale=.55]{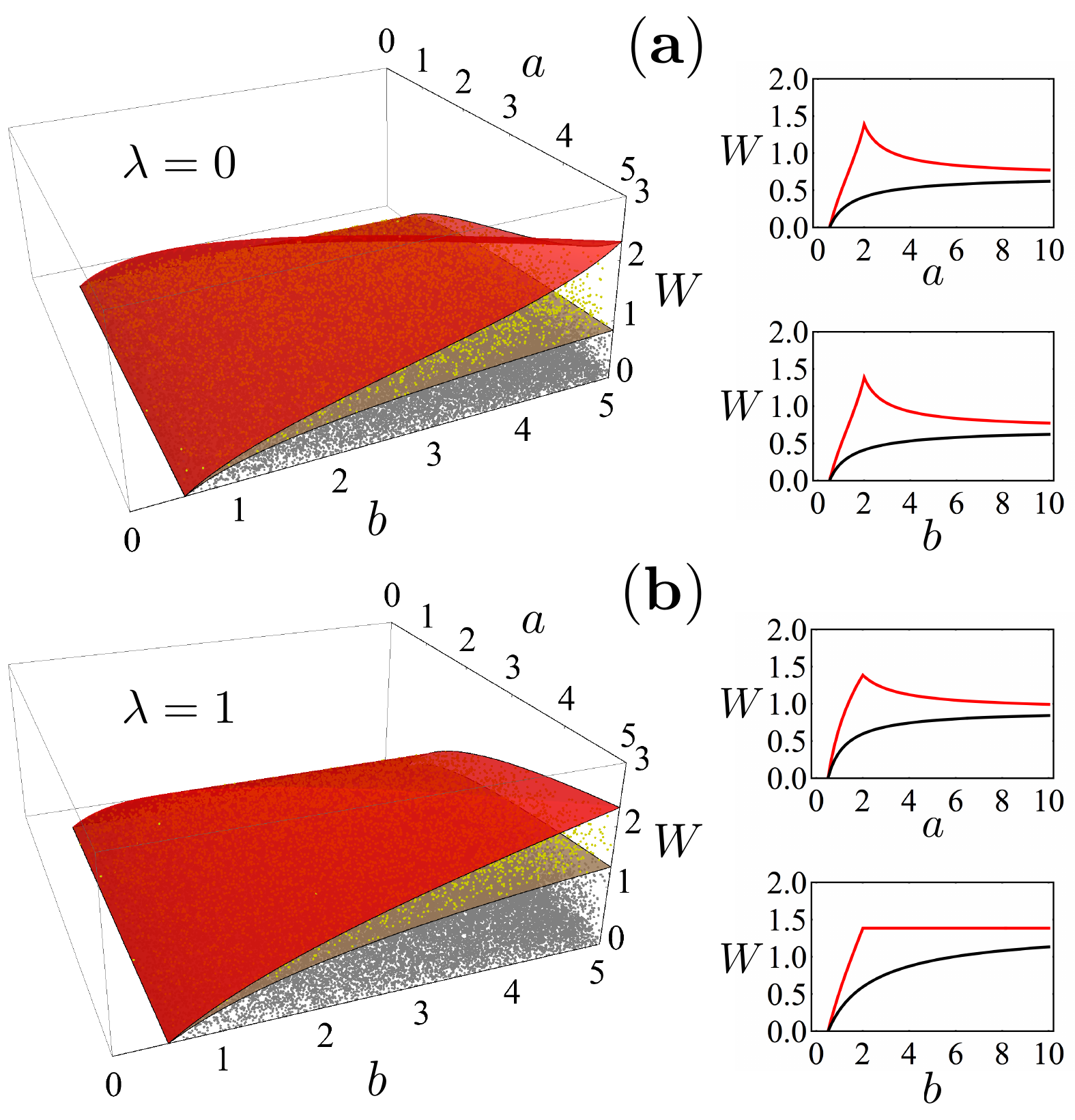}
\caption{Extractable work $W$ (in units of $k_BT$) against local energies $a$ and $b$ for randomly generated STSs. Each point 
corresponds to a state obtained by a uniform sampling of the parameters $a$, $b$ and $c$. Points corresponding to entangled 
(separable) states are marked in yellow (gray). Panel ({\bf a}) refers to homodyne detection ($\lambda=0$), while panel ({\bf b}) 
refers to heterodyne detection ($\lambda=1$). Maximum and separable work $W_{\text{max}}^{(k)}$ and $W_{\text{sep}}^{(k)}$, $k=0,1$
correspond to red and gray surfaces, respectively. Finally,  on the right column sections of both plots are shown.
\label{f:PlotWorkSTS}}
\end{figure}

The very same analysis can be extended to the class of non-symmetric squeezed thermal states (STSs) having different thermal 
occupation in each mode, obtained by setting $c_{ab}=\text{diag}(c,-c)$ in Eq.~\eqref{StandardForm}. 
The parameters fullfil $a\ge \frac12$, $b\ge \frac12$ and $\vert c\vert \le \max\left\{\sqrt{\bigl(a+\frac{1}{2}\bigr)
\bigl(b-\frac{1}{2}\bigr)},\sqrt{\bigl(a-\frac{1}{2}\bigr)\bigl(b+\frac{1}{2}\bigr)}\right\}$. The extractable work $W^{(\lambda)}$ in this case reads
\begin{equation}\label{WorkSTS}
W^{(\lambda)}=\frac{1}{2} \sum_{k=0,1}\ln \left[\frac{a (2 b\lambda^k +\lambda^{1-k})}{2( a b-c^2) \lambda^k+a\lambda^{1-k}}\right] \, ,
\end{equation} 
which still does not depend on the measurement angle and reduces to Eq.~\eqref{Wsymm} when $b\rightarrow a$.  
Also here one can verify that, for fixed $a$ and $b$, by increasing  $c$, one  both increases the value of $W^{(\lambda)}$ and moves from the class of 
separable states to entangled states (or increases the entanglement). Thus the extractable work, supplemented with the local purities, still provides a 
necessary and sufficient condition for the entanglement of the initial state, by checking the condition $W^{(\lambda)} > W_{\text{sep}}^{(\lambda)}$, where 
$W_{\text{sep}}^{(\lambda)}$ is obtained by substituting $c_\text{sep} = \sqrt{(a-1/2)(b-1/2)}$. In Fig.~\ref{f:PlotWorkSTS} we show the most relevant cases 
of homodyne/heterodyne detection, along with the separability thresholds $W_{\text{sep}}^{(\lambda)}$ and maximum work $W_\text{sep}$, whence we can 
see that for $\lambda=1$ the extractable work is no longer symmetric with respect to $a$ and $b$, and $W_{\text{max}}$ now acquires a dependence on the 
measurement. We also stress that the maximum amount of work extractable out of a separable state is achieved by a heterodyne measurement, i.e. not
by a projective measurement.

\par
\begin{figure}[t!] 
\includegraphics[scale=.6]{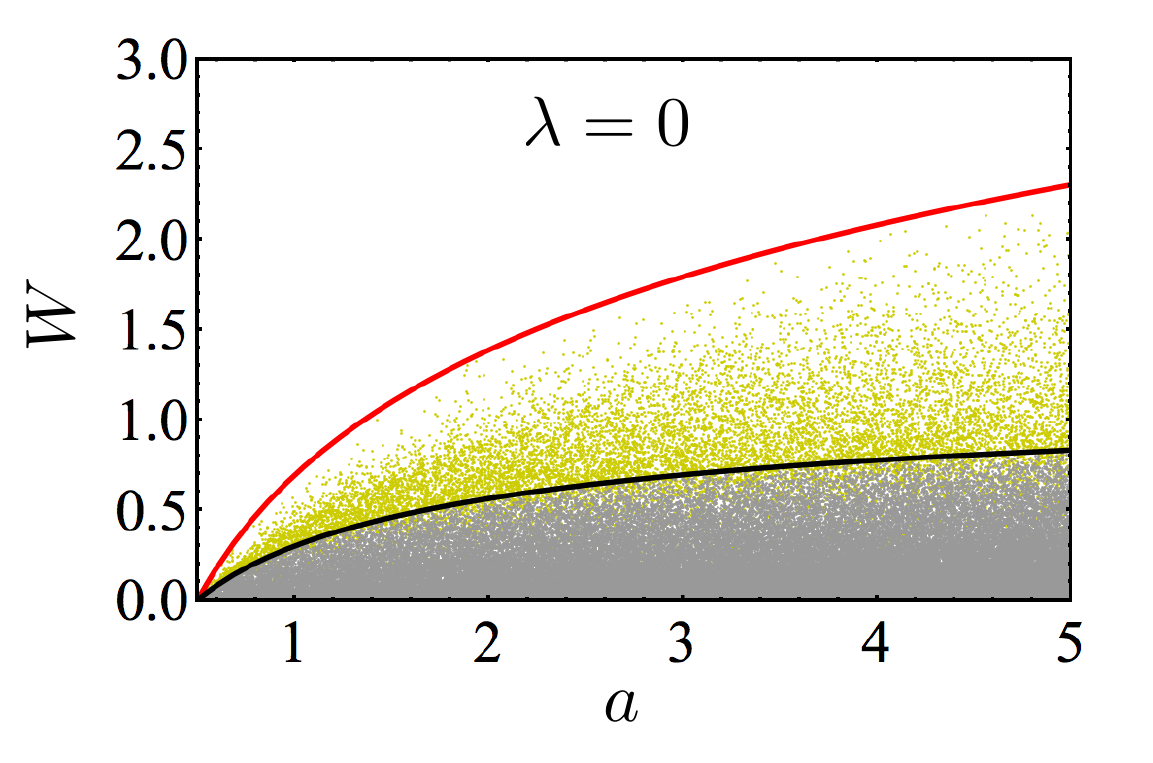}
\caption{Extractable work $W$ (in units of $k_BT$) for a STS against the parameter $a$. Random generated states are constrained to have
$b \le b_{\text{max}}$ where we set  $b_{\text{max}}=3$. Points corresponding to entangled (separable) states are marked in yellow (gray)
and we performed homodyne detection. The black curve is given by $W_{\text{sep}}^{(0)}$ evaluated at $b=b_{\text{max}}$, while the red one by
$W_{\text{max}}^{(0)}$ evaluated at $b=a$.
\label{f:PlotWorkLimited}}
\end{figure}

We can then consider the case where exhaustive information about the local purities is not available. Let us assume that only one local energy is 
known exactly, say $a$, while on the other only an upper bound is available, i.e. $b \le b_{\text{max}}$. This situation is illustrated in Fig.~\ref{f:PlotWorkLimited} 
for the case of a homodyne measurement.  Since (gray) points corresponding to separable states only occupy the portion of the graph below a threshold, we can 
conclude that the criterion is still sufficient for entanglement detection. The separability threshold is provided by the corresponding expression of the STS 
$W_{\text{sep}}^{(\lambda)}$  evaluated at $b=b_{\text{max}}$, while $W_{\text{max}}^{(\lambda)}$ is evaluated along the bisection line $b=a$.

\subsection{General two-mode Gaussian state}
\begin{figure}[h!] 
\includegraphics[scale=.55]{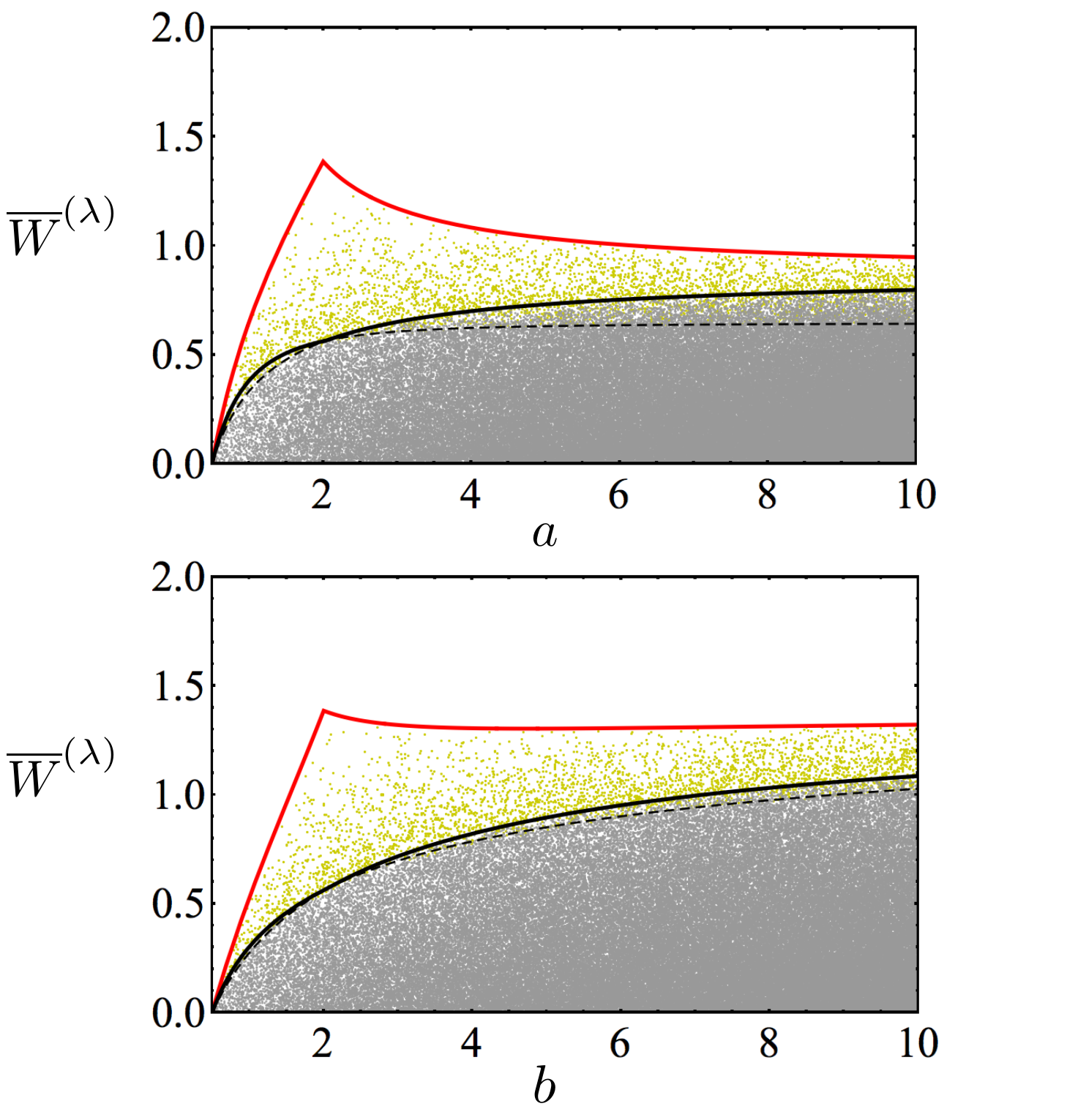}
\caption{Extractable work $\overline{W}^{(\lambda)}$ (in units of $k_BT$) averaged over the detection angle $\phi$ against the local energies $a$ and $b$. Points 
are obtained by random sampling. Detection strength has been fixed to a generic value $\lambda=3$. 
Points corresponding to entangled (separable) states are marked in yellow (gray). Maximum and separable work $W_{\text{max}}^{(\lambda)}$ and 
$\overline{W}_{\text{sep}}^{(\lambda)}$ correspond to red and black curves, respectively.
\label{f:PlotWorkAsymm}}
\end{figure}

Let us now consider two-mode states in standard form Eq.~\eqref{StandardForm}. 
In this  case the expression of the extractable work $W^{(\lambda)}(\phi)$ depends on the measurement angle, so that we will consider the average 
$\overline{W}^{(\lambda)}=\frac{1}{2\pi}\int_0^{2\pi}\mathrm{d}\phi\, W^{(\lambda)}(\phi)$. 
In this case we cannot prove any analytical relation between the extractable work and the separability of the initial bipartite state. In Fig.~\ref{f:PlotWorkAsymm} 
we display $\overline{W}^{(\lambda)}$ against the local energies for randomly generated states of the form~\eqref{StandardForm}, and we observe that the amount 
of work extractable from separable states (grey points) looks upper bounded, and thus seems to provide a necessary condition for detecting entanglement. 
Numerical inspection shows that the correlations, and in turn the extractable work $\overline{W}^{(\lambda)}$, is maximized, at fixed $a$, $b$ and $c$, by either the 
corresponding STS (recovered in the limit $d\rightarrow-c$) for which we already know the bound, or by states having a covariance matrix given by 
Eq.~\eqref{StandardForm} with $c_{a,b}=\text{diag}(c,0)$.
We denote members of the latter class by $\sigma'$.
These states are always separable, but sometimes they can be more correlated than a separable STS (with same $a$ and $b$). For these states the 
bounds on physicality and separability coincide. We will refer to that bound, to be averaged over $\phi$, as ${W}_{\text{sep}}^{(\lambda)}(\sigma')$, and
the corresponding  analytical expression is reported in Appendix B. Therefore we propose the following upper bound on the extractable work from separable states 
\begin{equation}\label{UpperSep}
\overline{W}_{\text{sep}}^{(\lambda)}(\sigma_{ab}) =\max\left[{W_{\text{sep}}^{(\lambda)}}(\sigma_{\text{STS}}),\overline{W}_{\text{sep}}^{(\lambda)}(\sigma')\right] \, .
\end{equation}

In Fig.~\ref{f:PlotWorkAsymm}, $\overline{W}_{\text{sep}}^{(\lambda)}$ is shown in black, with the dotted curve showing the smaller of the two components 
appearing in Eq.~\eqref{UpperSep}. We can see that for small $a$ and $b$, states $\sigma'$ result in more extractable work than $\sigma_{\text{STS}}$, and that we 
cannot find any random separable state violating the bound. This result is in agreement with the findings of Ref~\cite{WorkVedral}. Our result holds for generic measurement 
strength $\lambda$ and is not restricted to projective measurements ($\lambda\rightarrow 0,\infty$). 

\section{Measurement on both parties}\label{s:TwoMeas}
In this section we address the second scenario, sketched in Fig.~\ref{f:Protocol} (\textbf{c}) and addressed for qubits in Ref.~\cite{WorkVedral}, where  
both the demons Alice and Bob perform measurements on their reduced state. 
The second Gaussian measurement performed by Alice is described as $\hat\pi_a(Y)=\pi^{-1}\hat D_a (Y)\hat{\varrho}^{\pi_a}\hat D_a^{\dagger}(Y)$ 
where $\hat D_a(Y)=\exp(Y \hat a^{\dagger}-Y^* \hat a)$  and $\hat{\varrho}^{\pi_a}$ is a pure Gaussian state with covariance matrix 
$\gamma^{\pi_a}=R(\theta)\text{diag}(\mu/2,\mu^{-1}/2)R(\theta)^T$, $\mu \in[0,\infty]$.  The probability distribution corresponding to the measurement 
on mode $\hat{a}$, conditioned by the measurement $\hat\pi_b(X)$ performed on mode $\hat{b}$, turns out to be a Gaussian distribution whose covariance 
matrix is independent on the outcome of the measurements, i.e. $\sigma_{ab}^{\pi_b, \pi_a}=\sigma_{a}^{\pi_b}+\gamma^{\pi_a}$, where 
$\sigma_{a}^{\pi_b}$ is given by Eq.~\eqref{ReducedState}. 
Since work is extracted by a diffusion-like process in the phase space, starting with a less localized state intuitively results in less work extracted. In this case 
the extractable work is quantified via the Shannon entropies of the corresponding probability distribution  $H(\text{Pr}(X,Y))$ which is equal to the entropy of 
the Gaussian distribution $H(\sigma_{ab}^{\pi_b, \pi_a})$. We thus have
\begin{align}\label{WorkTwoMeas}
W&=k_B T \left[ H(\sigma_a+\gamma^{\pi_a})-H(\sigma_{ab}^{\pi_b, \pi_a})\right] \nonumber \\
&=\frac{k_B T}{2}\ln \left[\frac{\det (\sigma_a+\gamma^{\pi_a})}{\det(\sigma_{a}^{\pi_b}+\gamma^{\pi_a})}\right] \, .
\end{align}
The generic expression $W=W^{(\lambda,\mu)}(\phi,\theta)$ must then be averaged over the angles $\theta,\,\phi$. This is the work extracted from the statistics of the outcome 
distributed according to a Gaussian distribution with covariance matrix $\sigma_{ab}^{\pi_b, \pi_a}$. Expression~\eqref{WorkTwoMeas} also elucidates why we chose the
R\'enyi-2 entropy Eq.~\eqref{ShanWig} in place of the usual Von Neumann entropy as the entropic quantifier for a  \textit{state}. With that choice the one- and two-measurement
work extracting protocols are ``smoothly linked'' since the respective work outputs Eqs.~\eqref{WorkTwoMeas} and~\eqref{WorkExtRenyi} are related by a Gaussian convolution.
\begin{figure}[t!] 
\includegraphics[scale=.6]{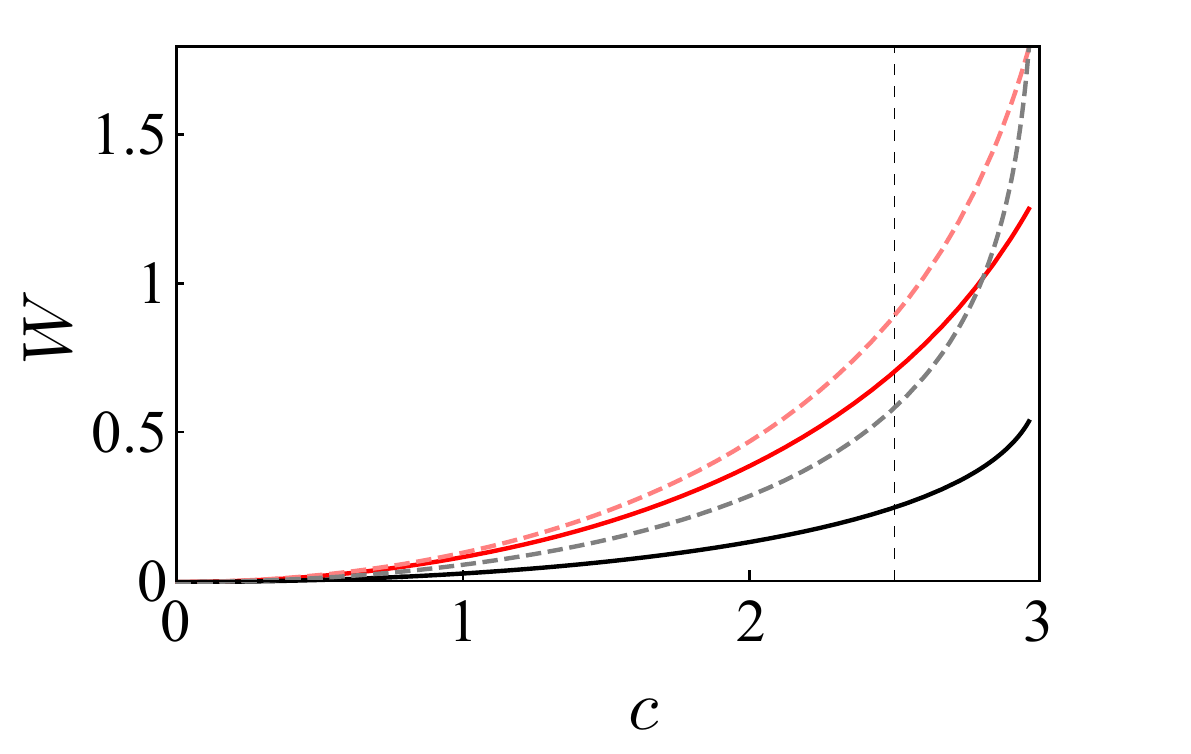}
\caption{Extractable work $W$ (in units of $k_BT$) for $\sigma_{\text{symm}}$ against the parameter $c$ for fixed $a=3$. 
The red curve is for $W^{(1,1)}$ while the black for $\overline{W}^{(0,0)}$. We also show a comparison with work extracted 
via single heterodyne detection $W^{(1)}$ (red dashed) homodyne detection $W^{(0)}$ (black dashed).  
The vertical dashed lines refers to the value $c_{\text{sep}}=a-1/2$. 
\label{f:PlotWork2}}
\end{figure}

For the case of a symmetric STS and two homodyne/heterodyne measurements we get
\begin{equation}
W^{(1,1)}=\frac{1}{2} \ln \left[\frac{(2 a+1)^4}{\left((2 a+1)^2-4 c^2\right)^2}\right]
\end{equation}
and
\begin{equation}
W^{(0,0)}(\phi,\theta)=\frac{1}{2} \ln \left[\frac{2 a^2}{2 a^2-c^2[ \cos (2 (\theta +\phi ))+1]}\right] \, .
\end{equation}
From the last expression we see that for $\theta+\phi=(2k+1)\pi/2$, $k\in \mathbb{Z}$ the extractable work identically vanishes, which
explains why the meaningful quantity is given by $\overline{W}^{(0,0)}$.
In Fig.~\ref{f:PlotWork2} we compare $\overline{W}^{(0,0)}$ ($W^{(1,1)}$) to $W^{(0)}$  ($W^{(1)}$). We can see the reduction of the 
extractable work due to the smearing of the distribution imparted by the second measurement. Contrary to the single-measurement scenario, 
now also for a two-mode squeezed vacuum $c=c_{\text{max}}$ a considerable gap between $\overline{W}^{(0,0)}$  and $W^{(1,1)}$ opens,
which significantly penalizes homodyne measurements.

\subsection{Relation with the mutual information}
Interestingly, when Alice and Bob both perform heterodyne detection, a clear connection between the extractable work $W^{(1,1)}$ and a form 
of mutual information emerges. The extractable work $W^{(1,1)}$ can be cast in the form 
\begin{equation}\label{Wehrl}
W^{(1,1)}=\frac{k_B T}{2} \ln \left(\frac{\tilde I_1 \tilde I_2}{\tilde I_4}\right)
\end{equation}
where $\tilde I_{1(2)}=\det{\tilde \sigma_{a(b)}}$ and $\tilde I_{4}=\det{\tilde \sigma_{ab}}$ are the symplectic invariants of the covariance matrix 
$\tilde \sigma_{ab}=\sigma_{ab}+\mathbb{1}/2$. The latter can be seen as the result of a convolution between the original covariance matrix and 
the vacuum. Indeed, it can be checked that Eq.~\eqref{Wehrl} equals $k_B T$ times the mutual information computed with the Wehrl entropy 
$S(\varrho)=-\int \text{d}\alpha Q(\alpha)\log Q(\alpha)$, i.e. the Shannon entropy of the Husimi Q-function $Q(\alpha)=\frac{1}{\pi} 
\langle\alpha\vert\varrho\vert\alpha\rangle$. The Husimi Q-function is related to the Wigner function through convolution with the vacuum.
\par
On the other hand, if we consider the case where Bob performs two sets of homodyne measurements at $\phi=0$ and $\phi=\pi/2$ (namely the
$q$-quadrature and $p$-quadrature), the work that Alice can extract can be expressed as
\begin{equation}\label{RenyiMutual}
W^{(0)}(q,p)=\frac{k_B T}{2} \ln \left(\frac{I_1 I_2}{I_4}\right) \, ,
\end{equation}
where $I_{1,2}$ and $I_4$ are now the local and global symplectic invariants of $\sigma_{ab}$. Eq.~\eqref{RenyiMutual} coincides with 
the mutual information computed with the R\'enyi-2 entropy $\mathcal{I}(\sigma_{a:b})=S_2(\sigma_{ab}\vert\vert\sigma_a \oplus \sigma_b)$, 
namely the Kullbac-Leibler divergence between the joint Wigner function and the product of the reduced ones~\cite{Adesso}.
It can also be checked that a second potential measurement performed by Alice is inconsequential.  

\section{Work extraction beyond the Gaussian framework}\label{s:BeyondGauss}
So far we have assumed the demons shared a Gaussian state (in standard form) and implemented Gaussian measurements. In particular this entails that the 
reduced states of Alice and Bob are both thermal, so that the amount of work extracted (by Alice) is a direct measure of the one-way classical correlations
$\mathcal{J}^{\leftarrow}(\varrho_{ab})$. Let us here briefly explain how the expression of the extractable work can be generalized to a generic bipartite state. 
We will extend the notation adopted for covariance matrices $\sigma_{ab}$ to density operators $\varrho_{ab}$, and measure the entropy by the Von Neumann 
entropy $S(\varrho_{ab})=-\tr{\varrho_{ab} \ln \varrho_{ab}}$. Bob performs a measurement $\hat\pi_b(X)\ge0$, $\int \text{d}X\hat\pi_b(X)=\mathbb{1}$ on his 
side getting the $X$-outcome with probability $p(X)=\tr{\varrho_{ab}\mathbb{1}_a\otimes \hat\pi_b(X)}$, and Alice's reduced state must be updated to 
$\varrho_{a\vert X}$. The (non optimized) one-way classical correlations is given by $\mathcal{J}^{\leftarrow}(\varrho_{ab})=S(\varrho_a)-\int \text{d}X p(X) 
S(\varrho_{a\vert X})$. Recalling that $\varrho_a^{\text{eq}}=Z_a^{-1}\exp(-H_a/k_B T)$ is the final equilibrium state after thermalization with the reservoir, 
and comparing with Eq.~\eqref{WorkExt}, in general we have
\begin{equation}
W=k_B T \left[ \mathcal{J}^{\leftarrow}(\varrho_{ab})+S(\varrho_a^{\text{eq}})-S(\varrho_a)\right]\, . 
\end{equation} 
We notice that $\mathcal{J}^{\leftarrow}(\varrho_{ab})\ge 0$ and $S(\varrho_a^{\text{eq}})\ge S(\varrho_a)$, being $\varrho_a^{\text{eq}}$ the equilibrium state with
the \textit{same average energy}, so that the presence of initial quantum coherence in Alice state leads to and increased amount of extractable work. 
By adding and subtracting the term $k_B T\, \tr{\varrho_a \ln \varrho_a^{\text{eq}}}$ we can rewrite the previous equation as
\begin{equation}\label{WorkBeyond}
W=k_B T \left[ \mathcal{J}^{\leftarrow}(\varrho_{ab})+S(\varrho_a\vert\vert \varrho_a^{\text{eq}})\right]+\Delta Q_a \, , 
\end{equation} 
where $S(\varrho\vert\vert\sigma)=\tr{\varrho\ln\varrho-\varrho\ln\sigma}$ is the quantum relative entropy between two states~\cite{VlatkoRevEntropy} 
and $\Delta Q_a=\tr{(\varrho_a^{\text{eq}}-\varrho_a)H_a}$ is the heat absorbed from the bath in an isothermal expansion from the pre-measurement state
to the final state. When $\varrho_a=\varrho_a^{\text{eq}}$, as for Gaussian states in standard form, both the extra terms in Eq.~\eqref{WorkBeyond} vanish and the 
previous result is recovered. However, we notice that Eq.~\eqref{WorkBeyond} also applies to Gaussian states with non-diagonal reduced covariance matrix 
$\sigma_a\neq\text{diag}(a,a)$. 
We plan to further explore this relation in future works. 

\section{Work extraction from tripartite Gaussian states}\label{s:Tripartite}

We now move to investigate the extraction of work from a multipartite system. In analogy with the bipartite case, one can think of the extracting protocol 
as a continuous-variable Szilard engine with a multipartite working substance and a demon acting on each party. As in the previous sections, we are interested 
in the extractable work as a tool to investigate the nature of the correlations shared within the working medium. 
The classification of entanglement in multipartite systems is an extremely challenging problem~\cite{EntanglementRev}. 
In the following we will focus on the tripartite case, whose (in)separability structure is already considerably richer and more complex than the bipartite case. 
Let us consider a tripartite Gaussian state with covariance matrix 
\begin{equation}\label{sigmatri}
\sigma_{abc}=
\left(
\begin{array}{ccc}
\sigma_a  & c_{ab} & c_{ac} \\[1ex]
c_{ab}^T &  \sigma_b & c_{bc} \\[1ex]
c_{ac}^T  & c_{bc}^T&  \sigma_c
\end{array}
\right) \, ,
\end{equation}
where $\sigma_j$ is the reduced covariance matrix of each mode and $c_{jk}$ contains the correlations between modes $j$ and $k$, where 
$j, k\in\{a,b,c\}$, $j\neq k$. When considering a given bipartition of the state, say $(a b,c)$, we can equivalently employ the following notation 
\begin{equation}\label{sigmabip}
\sigma_{abc}=
\left(
\begin{array}{cc}
\sigma_{ab} \, & c_{ab,c} \\[1ex]
c_{ab,c}^T &  \sigma_c
\end{array}
\right) \, ,
\end{equation}
where $c_{ab,c}=\bigl(c_{ac} \; c_{bc}\bigr)^T$ is a $4 \times 2$ matrix containing correlations between $c$ and the two-mode state $ab$. 
\par
Let us recall the separability structure of the class Eq.~\eqref{sigmatri}. For any bipartition of the state, positivity under partial transposition (PPT) 
provides a necessary and sufficient condition for separability. The PPT criterion singles out four distinct (in)separability classes: (i) states which are not 
separable under any bipartition of the modes. These states are called fully inseparable and  share genuine tripartite entanglement. (ii) States which 
are separable with respect to one bipartition only, referred as 1-biseparable states. (iii)  States which are separable for two different bipartitions (2-biseparable states), 
and (iv) states separable under all the three bipartitions (3-biseparable states)~\cite{GiedkeTri}. Notice that fully separable states of thee three modes, i.e. states of the form
$\hat\varrho_{\text{sep}}=\sum_k p_k \hat\varrho_{a,k}\otimes\hat\varrho_{b,k}\otimes\hat\varrho_{c,k}$ belong to class (iv). 
Thus entangled states are present in all the classes listed, ranging from genuinely tripartite entangled states in (i) to bound entangled states in (iv).
\par  
In the natural extension of the work extracting protocol that we consider, Alice extracts work from a local heat bath by acting on her state, after Bob and Charlie
performed local measurements on their modes. Again, by restricting to Gaussian measurement with pure seed, the conditional state of  Alice and Bob 
after Charlie's measurement $\hat \pi_c$ is given by
\begin{align}
\sigma_{ab}^{\pi_c}&=\sigma_{ab}-c_{ab,c}(\sigma_c+\gamma^{\pi_c})^{-1}c_{ab,c}^T \\
&= \left(
\begin{array}{cc}
\sigma_a^{\pi_c} \, & c_{ab}^{\pi_c} \\[1ex]
c_{ab}^{\pi_c,T} &  \sigma_b^{\pi_c}
\end{array}
\right) \, ,
\end{align}
so that a second measurement $\hat \pi_b$ on Bob's side leaves Alice with the conditional state
\begin{equation}\label{CondStateTri}
\sigma_{a}^{\pi_b, \pi_c}=\sigma_{ab}-
\left(
\begin{array}{c}
 c_{ac} \\
 c_{bc} 
\end{array}\right)
(\sigma_c+\gamma^{\pi_c})^{-1}\bigl(c_{ac} \; c_{bc}\bigr)\, .
\end{equation}
By letting the state Eq.~\eqref{CondStateTri} thermalize, Alice can thus extract an 
amount of work given by
\begin{equation}\label{WorkExtTri}
W=\frac{k_B T}{2}\ln \left(\frac{\det \sigma_a}{\det\sigma_{a}^{\pi_b, \pi_c}}\right) \, .
\end{equation}
We are now in position to address how the different separability classes of states Eq.~\eqref{sigmatri} affect the work extracting protocol. 

\subsection{Tripartite pure states}
Let us first address the case of pure tripartite states $\sigma_{abc}^P$. 
For these states an explicit parametrization can be given in terms of the diagonal elements alone. The standard
form of a pure tripartite state is given by 
\begin{equation}\label{SigmaPureTri}
\sigma_{abc}^P=\left(
\begin{array}{cccccc}
 a & 0 & c_{ab}^+ & 0 & c_{ac}^+ & 0 \\
 0 & a & 0 & c_{ab}^- & 0 & c_{ac}^- \\
  c_{ab}^+ & 0 & b & 0 & c_{bc}^+ & 0 \\
 0 & c_{ab}^- & 0 & b & 0 & c_{bc}^- \\
 c_{ac}^+ & 0 & c_{bc}^+ & 0 & c & 0 \\
 0 & c_{ac}^- & 0 & c_{bc}^- & 0 & c \\
\end{array}
\right) \, ,
\end{equation}
where $a, b, c \ge 1/2$ and
\begin{widetext}
\begin{equation}
c_{ij}^{\pm}=\frac{\sqrt{\left[4 (i-j)^2-(2 k-1)^2\right] \left[4 (i-j)^2-(2 k+1)^2\right]} \pm \sqrt{\left[4 (i+j)^2-(2 k-1)^2\right] \left[4 (i+j)^2-(2 k+1)^2\right]}}{16 \sqrt{i j}} \,,
\end{equation}
\end{widetext}
with $j, k\in\{a,b,c\}$, $j\neq k$. For such states, the separability structure is considerably simpler: if $a=b=c=1/2$ the state is the product of three single-mode vacua, 
otherwise it can either be the product of the vacuum in one mode and a maximally entangled state of the other two
(1-biseparable state), or be fully inseparable. In particular, no states in class (iii) can be found, as well any state in (iv) which is not in the factorized form.
\par
It is well known that for the qubit case two inequivalent classes of pure tripartite entangled states emerge: GHZ states with maximal genuine tripartite entanglement 
and zero bipartite entanglement in any two-qubit reduction, and W states with maximal bipartite entanglement across any bipartition but vanishing tripartite entanglement
~\cite{GHZ/W}. On the contrary, it can be shown that the subclass of pure symmetric Gaussian states retains at the same time the entanglement properties of both GHZ 
and W states. By either maximizing the bipartite entanglement in any two mode reduction (W-like) or maximizing the genuinely tripartite entanglement (GHZ-like), the 
same family of state is singled out, namely state of the form Eq.~\eqref{SigmaPureTri} with $a=b=c$ and 
\begin{equation}
c^{\pm}=\frac{4a^2-1\pm\sqrt{(4a^2-1)(36a^2-1)}}{16a} \, .
\end{equation}
Let us recall that in Ref.~\cite{WorkVedralTripartite} a suitable strategy based on the local extractable work was proposed to distinguish between GHZ and W states. 
\par
In the present case the expression for the extractable work Eq.~\eqref{WorkExtTri} can be analytically evaluated and reads
\begin{equation}\label{WP}
W^P=\ln 2a.
\end{equation}   
By  direct comparison with Eq.~\eqref{WMaxSymm} we see that $W^P$ coincides with the maximum work extractable from a two mode symmetric state.
This amount of work turns out to be independent on the measurements implemented by the demons Bob and Charlie. 
Moreover, the same amount of work as in Eq.~\eqref{WP} can be extracted from the  tripartite pure state in Eq.~\eqref{SigmaPureTri} for fixed purities $b$ and $c$,
independently on the measurement.  
Therefore, since the same amount of work is extracted from a pure bipartite entangled state and a fully inseparable three-mode sate (with same $a$), we conclude that 
the demon Alice would not boost work extraction by entangling her Gaussian Szilard engine with a third mode.      
On the other hand, based on the amount of work extracted, Alice cannot distinguish between states belonging to class (ii) of the form $\ket{0}_b\otimes \hat S_2(\xi)\ket{00}_{ac}$ 
and $\ket{0}_c\otimes \hat S_2(\xi)\ket{00}_{ab}$ and a genuinely tripartite symmetric state.
This fact seems to point out that bipartite entanglement is the essential resource behind the work extracting protocol. If the Gaussian demon Alice had to decide based on the 
extractable work only, she could not tell whether she is extracting work from a bipartite Szilard engine with a pure entangled working substance or from a tripartite one.

\subsection{Symmetric mixed states}
\begin{figure}[t!] 
\includegraphics[scale=.51]{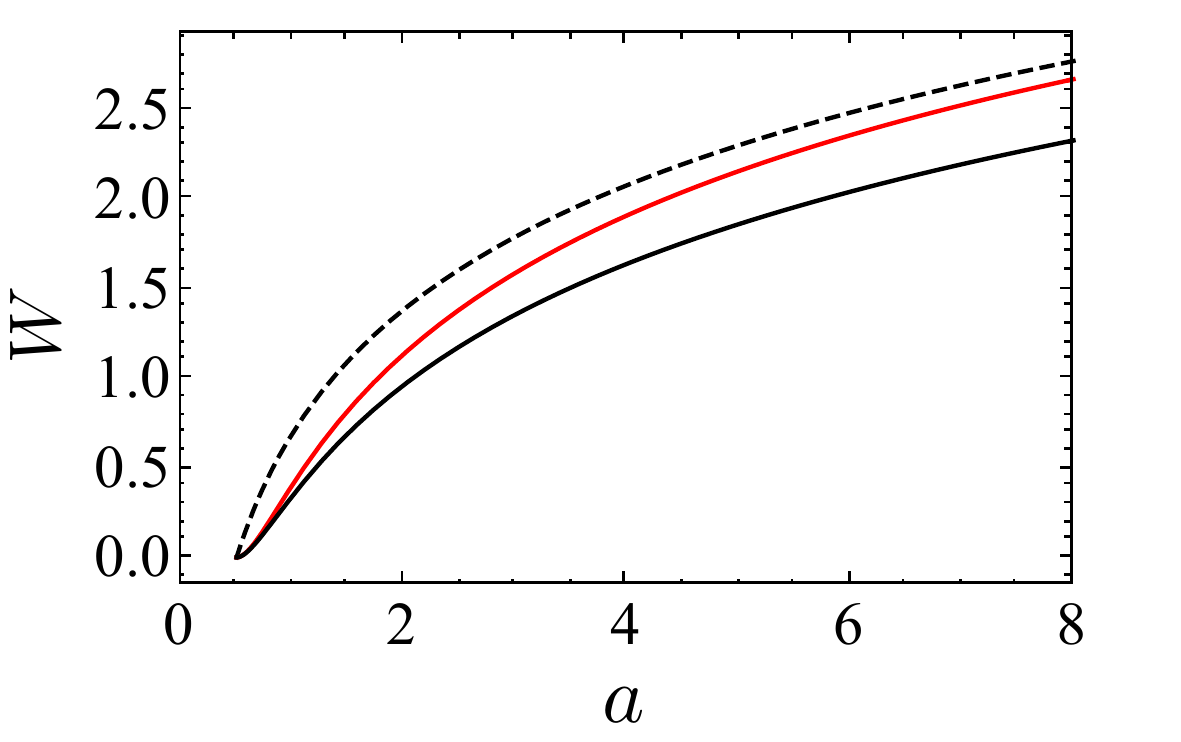}
\caption{Extractable work $W$ (in units of $k_BT$) against local energy $a$ for a fully symmetric mixed state $\sigma_{abc}^S$ when 
either heterodyne detection (red curve) or homodyne detection (black curve) is performed on both Bob's and Charlie's side. In case of
homodyne detection the work has been averaged over the two angular variable. The black dashed line corresponds to the work
extracted from a pure symmetric tripartite state $\sigma_{abc}^P$.
\label{f:PlotWorkTriSymm}}
\end{figure}

Another relevant class is the one of fully symmetric mixed states, namely tripartite states invariant under the permutation of any mode. 
Their covariance matrix in block form is given by~\cite{GerardoTri2006} 
\begin{equation}\label{sigmatrisymm}
\sigma_{abc}^S=
\left(
\begin{array}{ccc}
\sigma_a  & C & C \\[1ex]
C^T &  \sigma_a & C \\[1ex]
C^T  & C^T&  \sigma_a
\end{array}
\right) \, ,
\end{equation}
with $\sigma_a=a\mathbb{1}_2$ and $C=\mathrm{diag}(c^+,c^-)$ with elements 
\begin{align}
c^+&=\frac{4a^2-5+\sqrt{36a^2(4a^2-2)+25}}{16a} \, , \nonumber \\
c^-&=\frac{5-36a^2+\sqrt{36a^2(4a^2-2)+25}}{48a} \, .
\end{align}
These are states which are either factorized for $a=1/2$ or fully inseparable whenever $a>1/2$. They can be obtained by 
maximizing the entanglement between any bipartition, while at the same time imposing no entanglement to be present within the 
two-mode state. For such a class of states the extractable work can be computed analytically, although the resulting expressions are 
quite involved. In Fig.~\ref{f:PlotWorkTriSymm} we plot the extractable work for states $\sigma_{abc}^S$ when demons Bob and 
Charlie perform either heterodyne detection (red curve) or homodyne detection (black curve). The dashed curve represents $W^P$,
i.e. the work extracted with pure tripartite states  $\sigma_{abc}^P$ for any measurement. In particular we notice that $W^P$ is 
asymptotically reached for heterodyne detection even if the state is mixed. We can thus conclude that, besides the presence of 
genuinely tripartite entanglement, also the global mixedness of the state does not play a crucial role when work extraction is 
concerned.   

\subsection{Tripartite mixed states}

\begin{figure}[t!] 
\includegraphics[scale=.51]{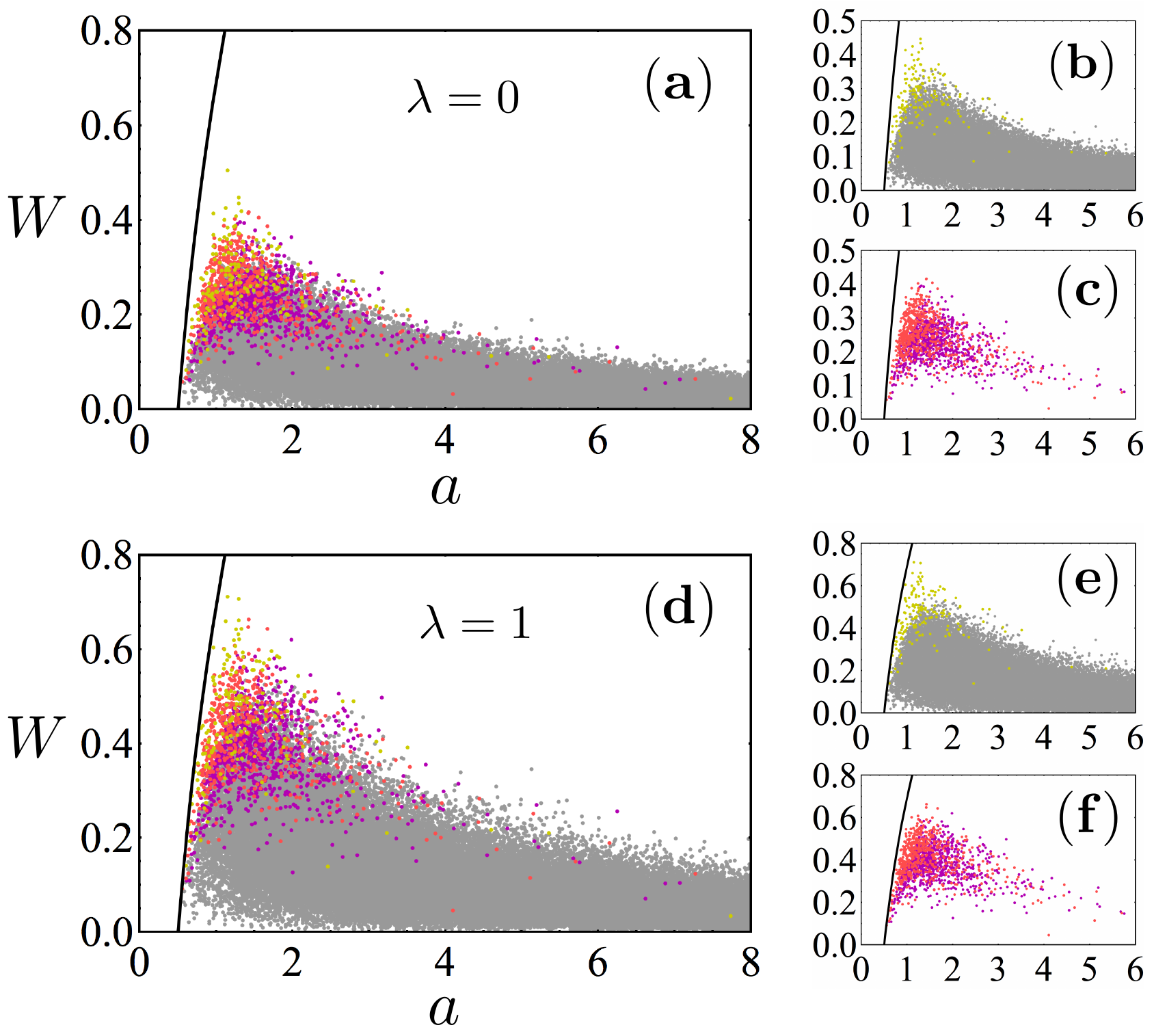}
\caption{Extractable work $W$ (in units of $k_BT$) against local energy $a$ for randomly generated tripartite states $\sigma_{abc}^M$. 
Points corresponding to fully inseparable states are marked in yellow, 1-biseparable in red, 2-biseparable in purple and 3-biseparable 
in grey. Panels ({\bf a}) - ({\bf c}) refer to homodyne detection ($\lambda=0$) on Bob's and Charlie's side, while panels ({\bf d}) - ({\bf f})
 to heterodyne detection ($\lambda=1$). The maximum work $W_{\text{max}}$ corresponds to the black curve.
\label{f:PlotWorkTri}}
\end{figure}

In order to better understand the interplay between purity, bipartite and genuinely tripartite quantum correlations and work extraction, we now 
turn our attention to the generic mixed tripartite case. When cast in standard form~\cite{MixedTriStd}, the covariance matrix of a general tripartite 
state reads
\begin{equation}
\sigma_{abc}^M=
\left(
\begin{array}{cccccc}
 a & 0 & c_1 & 0 & c_3 & c_5 \\
 0 & a & 0 & c_2 & 0 & c_4 \\
  c_1 & 0 & b & 0 & c_6 & c_8 \\
 0 & c_2 & 0 & b & c_9 & c_7 \\
 c_3 & 0 & c_6 & c_9 & c & 0 \\
 c_5 & c_4 & c_8 & c_7 & 0 & c \\
\end{array}
\right) \, .
\end{equation}
Given that the fully-fledge problem cannot be attacked analytically, we proceed by randomly generating states $\sigma_{abc}^M$ sampling each of the twelve 
parameters from a uniform distribution. Such states are then classified by applying the PPT criterion across every bipartitions: states belonging to classes (i)-(iv) 
are colored in yellow, red, purple and grey, respectively. We then compute the extractable work Eq.~\eqref{WorkExtTri} when the Gaussian demons Bob and Charlie
perform their measurements, and average the work over the detection angles. The result is shown in Fig.~\ref{f:PlotWorkTri} for the relevant cases of homodyne 
[panels ({\bf a}) - ({\bf c})] and heterodyne detection [panels ({\bf d}) - ({\bf f})].  

\par
From Fig.~\ref{f:PlotWorkTri}
We can see that between state belonging to class (iv), which are either classically correlated or possess bound entanglement at most, and genuine
tripartite entangled states (i) there is no a dramatic difference as far as work extraction is concerned, meaning that the overlapping region 
is significant. As reasonably expected, genuine tripartite entanglement on average leads to higher values of extracted work. However, the distributions
of the work values does not seem to be lower bounded. In panels ({\bf b}), ({\bf e}) we highlight this feature for the case of homodyne and heterodyne, respectively. 
\par 
Moreover, through work extraction the demons can hardly discriminate between the case where entanglement is present across all the three bipartitions (yellow 
points) or or just across one of them (purple points). This is another evidence that work extraction is not sensitive to entanglement being shared between 
two demons (either Alice and Bob or Alice and Charlie) rather than among all the three of them. In Fig.~\ref{f:PlotWorkTri} panels ({\bf c}), ({\bf f}) we 
highlight the work extracted from 1-biseparble and 2-biseparable states.
\par

\section{Conclusions and outlook}\label{s:Conclusions}
We have formulated a protocol for extracting work (out of a thermal bath) by means of a correlated quantum system subjected to measurements.
In particular, we focused on a fully Gaussian framework (Gaussian states and Gaussian measurements),
phrasing the protocol in terms of demons acting locally on a Szil\'ard engine with a multipartite working substance which may contain
quantum correlations. By exploiting the initial correlations and the measurement back-action, work can be extracted by one of the demons. 
We have then addressed the use of the work output as a detector of entanglement. We provided evidence that this the case for a  
two-mode Gaussian state in standard form. Moreover, for the subclass of squeezed thermal states we proved that the extractable work (together with 
the local purities) also provide a necessary condition for inseparability. 
Despite the focus on the Gaussian scenario, we showed how the  framework can be easily generalized to account for the presence of initial quantum 
coherence and generic measurement, thus going beyond the Gaussian framework. 
We then enquired wether the extractable work can be used to discriminate among richer inseparability structures, as the one provided by tripartite Gaussian 
states. We found that genuine tripartite entanglement can be hardly distinguished from bipartite one. 
In conclusion, sharing entanglement among many parties does not seem to boost the amount of work extracted by one of them and, conversely, the 
effectiveness of work-based separability criterion considerably weakens moving from two to three parties, even for a special class
of states such as the Gaussian one. 
\noindent
\section*{Acknowledgements}
This work was supported by the UK EPSRC (EP/L005026/1 and EP/J009776/1), the  Julian Schwinger Foundation 
(grant Nr. JSF-14-7-0000), the Royal Society Newton Mobility Grant (grant NI160057), the DfE-SFI Investigator Programme (grant 15/IA/2864)
and the EU Collaborative Project TherMiQ (Grant Agreement 618074). Part of this work was supported by COST Action 
MP1209 ``Thermodynamics in the quantum regime''. Marco Barbieri is supported by a Rita Levi-Montalcini fellowship of MIUR.
MGG acknowledges support from Marie Sk\l{}odowska-Curie Action H2020-MSCA-IF-2015 (project ConAQuMe, grant nr. 701154).

\section*{Appendix}
\subsection*{A. Expressions of the extractable work for squeezed thermal states}
Below we report the expressions of the extractable work for STSs, in the relevant case of homodyne ($\lambda=0$) and heterodyne ($\lambda=1$) detection
\begin{align}
W^{(0)}&=\frac{1}{2} \ln \left(\frac{a b}{a b-c^2}\right)\, , \\
W^{(1)}&=\frac{1}{2} \ln \left[\frac{(2 a b+a)^2}{\left(2 a b+a-2 c^2\right)^2}\right] \, .
\end{align}
In particular the the maximum extractable work and the upper bound on the work extractable from a separable state read
\begin{align}
W_{\text{max}}^{(0)}&=\frac{1}{2} \ln \left[\frac{4 a b}{1+2\vert a-b\vert}\right]\, , \\
W_{\text{max}}^{(1)}&=
\left\{
\begin{array}{cc}
\ln 2a  & \mathrm{if} \; \; a \le b  \\
 \\
\ln\left[\frac{2a(1+2b)}{1+4a-2b} \right] & \mathrm{otherwise} \, ,
\end{array}
\right.
\end{align}

\begin{align}
W_{\text{sep}}^{(0)}&=\frac{1}{2} \ln \left(\frac{4 a b}{2 a+2 b-1}\right)\, , \\
W_{\text{sep}}^{(1)}&=\frac{1}{2} \ln \left[\frac{4 (2 a b+a)^2}{(4 a+2 b-1)^2}\right] \, .
\end{align}

\subsection*{B. Expression of the separable work for two-mode Gaussian states in standard form}
For Gaussian states in standard form Eq.~\eqref{StandardForm} the bound on the work extractable out of a separable state is given by
$\overline{W}_{\text{sep}}^{(\lambda)}(\sigma_{ab}) =\max\left[{W_{\text{sep}}^{(\lambda)}}(\sigma_{\text{STS}}),\overline{W}_{\text{sep}}^{(\lambda)}(\sigma')\right]$,
where
\begin{equation}
W_{\text{sep}}^{(\lambda)}(\sigma_{\text{STS}})=\sum_{k=0,1} \ln \left[\frac{2 a(2 b\lambda^k +\lambda^{1-k})}{(2a+2b-1)\lambda^k+2a\lambda^{1-k}}\right]^{\frac12} \, ,
\end{equation}
and
\begin{widetext}
\begin{equation}
{W}_{\text{sep}}^{(\lambda)}(\sigma')=\frac12\ln\left(\frac{16 a^2 b (2 b+\lambda ) (2 b \lambda +1)}{\left(4 a^2-1\right) \left(4 b^2-1\right) \left(\lambda ^2-1\right) \cos(2\phi) 
+4 a^2 \left(4 b^2 \left(\lambda ^2+1\right)+8 b \lambda +\lambda ^2+1\right)+\left(4 b^2-1\right) \left(4 b \lambda +\lambda ^2+1\right)}\right) \, ,
\end{equation}
\end{widetext}
then to be averaged over $\phi\in[0,2\pi]$.

\end{document}